\documentclass[12pt,a4paper]{article}
\usepackage{amsfonts}
\usepackage{graphicx}
\usepackage{amsmath}
\usepackage{amssymb}
\usepackage{fancyhdr}

\input{tcilatex}

\begin{document}

\author{Piroska D\"{o}m\"{o}t\"{o}r
and Mih\'{a}ly G. Benedict\thanks{benedict@physx.u-szeged.hu} \\
{\it Department of Theoretical Physics, University of Szeged, Hungary} \\
}
\title{Coherent states and global entanglement in an $N$ qubit system}
\date{}
\maketitle

\begin{abstract}
We consider an $N$ qubit system and show that in the symmetric subspace, $%
\mathbb{S}$ a state is not globally entangled, iff it is a coherent state.
It is also proven that in the orthogonal complement $\mathbb{S}_{\bot}$ all
states are globally entangled.
\end{abstract}

\section{Nonentangled pure states}

A pure $N$ qubit state is not entangled by definition, if it is a product
state. In order to decide if a pure state is entangled or not, we shall use
the following formal method used first by Meyer and Wallach \cite{MW02},
who applied this procedure to define a measure of entanglement for $N$ qubit
states. The pure state $\left\vert \psi\right\rangle $ $\in$ $\mathbb{C}%
^{2^{N}}$, which is expanded in the standard basis $\{\left\vert
0\right\rangle ,\left\vert 1\right\rangle \}^{\otimes N},$ can be decomposed
for each $n=1,2\ldots N$ qubit as
\begin{equation}
\left\vert \psi\right\rangle =\left\vert 0\right\rangle _{n}\otimes\left\vert
u^{n}\right\rangle +\left\vert 1\right\rangle _{n}\otimes{\left\vert
v^{n}\right\rangle ,} \label{decomp}%
\end{equation}
where $\left\vert {u}^{n}\right\rangle $ and $\left\vert {v}^{n}\right\rangle
$ are vectors in $\mathbb{C}^{2^{N-1}}$ which are not normalized, in general.
Using the above decomposition (\ref{decomp}) one can show that $\left\vert
\psi\right\rangle $ is a product state if and only if $\left\vert
u^{n}\right\rangle $ is parallel to ${\left\vert v^{n}\right\rangle }$ for all
possible $n$.

First, assume that $\left\vert \psi\right\rangle $ is a product state and can
be written as
\begin{equation}
\left\vert \psi\right\rangle =%
{\displaystyle\bigotimes\limits_{i=1}^{N}}
(a_{i}\left\vert 1\right\rangle _{i}+b_{i}\left\vert 0\right\rangle _{i}),
\end{equation}
with $\left\vert a_{i}\right\vert ^{2}+\left\vert b_{i}\right\vert ^{2}=1$. In
this case $\left\vert u^{n}\right\rangle =b_{n}%
{\displaystyle\bigotimes\limits_{\substack{i=1\\i\neq n}}^{N}}
(a_{i}\left\vert 1\right\rangle _{i}+b_{i}\left\vert 0\right\rangle _{i})$,
while\linebreak${\left\vert v^{n}\right\rangle =}a_{n}%
{\displaystyle\bigotimes\limits_{\substack{i=1\\i\neq n}}^{N}}
(a_{i}\left\vert 1\right\rangle _{i}+b_{i}\left\vert 0\right\rangle _{i}),$
and it is obvious that these two vectors are parallel.

Second, let $\left\vert u^{n}\right\rangle $ be parallel to ${\left\vert
v^{n}\right\rangle }$ for all possible $n$: $\left\vert u^{n}\right\rangle
=\alpha_{n}\left\vert {v^{n}}\right\rangle $. Then $\left\vert \psi
\right\rangle $ can be written in the following form:%
\begin{equation}
\left\vert \psi\right\rangle =(1+\left\vert \alpha_{n}\right\vert ^{2}%
)^{-1/2}\left(  \left\vert 0\right\rangle _{n}+\alpha_{n}\left\vert
1\right\rangle _{n}\right)  \otimes\left\vert \tilde{u}^{n}\right\rangle
\ \quad\forall\ n, \label{parallel}%
\end{equation}
and the statement can be proven by induction. (Here the $N-1$ qubit state
$\left\vert \tilde{u}^{n}\right\rangle $ is normalized.) For $N=2$ it is
obviously true, because then\linebreak\ $\left\vert \psi\right\rangle
=(1+|\alpha_{1}|^{2})^{-1/2}\left(  \left\vert 0\right\rangle _{1}+\alpha
_{1}\left\vert 1\right\rangle _{1}\right)  \otimes\left\vert \tilde{u}%
^{1}\right\rangle $ and $\left\vert \tilde{u}^{1}\right\rangle $ is a one
qubit state.

Suppose now that the statement is true for $N-1$, and let's prove it for $N$.
We first use the decomposition (\ref{parallel}) with respect to the $i$-th
qubit, where $\left\vert \tilde{u}^{i}\right\rangle $ is now an $N-1$ qubit
state. Decompose $\left\vert \tilde{u}^{i}\right\rangle $ further, with
respect to the $j$-th qubit:
\begin{align}
\left\vert \psi\right\rangle  &  =\frac{\left(  \left\vert 0\right\rangle
_{i}+\alpha_{i}\left\vert 1\right\rangle _{i}\right)  }{\sqrt{1+\left\vert
\alpha_{i}\right\vert ^{2}}}\otimes\left\vert \tilde{u}^{i}\right\rangle
=\frac{\left(  \left\vert 0\right\rangle _{i}+\alpha_{i}\left\vert
1\right\rangle _{i}\right)  }{\sqrt{1+\left\vert \alpha_{i}\right\vert ^{2}}%
}\otimes(\left\vert 0\right\rangle _{j}\otimes\left\vert u^{ij}\right\rangle
+\left\vert 1\right\rangle _{j}\otimes{\left\vert v^{ij}\right\rangle
)=}\nonumber\\
&  {=}\left\vert 0\right\rangle _{j}\otimes\frac{\left(  \left\vert
0\right\rangle _{i}+\alpha_{i}\left\vert 1\right\rangle _{i}\right)  }%
{\sqrt{1+\left\vert \alpha_{i}\right\vert ^{2}}}\otimes\left\vert
u^{ij}\right\rangle +\left\vert 1\right\rangle _{j}\otimes\frac{\left(
\left\vert 0\right\rangle _{i}+\alpha_{i}\left\vert 1\right\rangle
_{i}\right)  }{\sqrt{1+\left\vert \alpha_{i}\right\vert ^{2}}}\otimes
{\left\vert v^{ij}\right\rangle ,}%
\end{align}
and compare this with
\begin{equation}
\left\vert \psi\right\rangle =\frac{\left(  \left\vert 0\right\rangle
_{j}+\alpha_{j}\left\vert 1\right\rangle _{j}\right)  }{\sqrt{1+\left\vert
\alpha_{j}\right\vert ^{2}}}\otimes\left\vert \tilde{u}^{j}\right\rangle .
\end{equation}
As a result we get
\begin{equation}
\frac{\left\vert \tilde{u}^{j}\right\rangle }{\sqrt{1+\left\vert \alpha
_{j}\right\vert ^{2}}}{=}\frac{\left(  \left\vert 0\right\rangle _{i}%
+\alpha_{i}\left\vert 1\right\rangle _{i}\right)  }{\sqrt{1+\left\vert
\alpha_{i}\right\vert ^{2}}}\otimes\left\vert u^{ij}\right\rangle
\ \ \ \ \ \ \alpha_{j}\frac{\left\vert \tilde{u}^{j}\right\rangle }%
{\sqrt{1+\left\vert \alpha_{j}\right\vert ^{2}}}=\frac{\left(  \left\vert
0\right\rangle _{i}+\alpha_{i}\left\vert 1\right\rangle _{i}\right)  }%
{\sqrt{1+\left\vert \alpha_{i}\right\vert ^{2}}}\otimes{\left\vert
v^{ij}\right\rangle }%
\end{equation}
which implies that $\alpha_{j}\left\vert u^{ij}\right\rangle =\left\vert
{v^{ij}}\right\rangle $. Then by hypothesis $\left\vert \tilde{u}%
^{i}\right\rangle $ can be written as a product state, and according to
$\left\vert \psi\right\rangle =\left(  1+\left\vert \alpha_{i}\right\vert
^{2}\right)  ^{-1/2}\left(  \left\vert 0\right\rangle _{i}+\alpha
_{i}\left\vert 1\right\rangle _{i}\right)  \otimes\left\vert \tilde{u}%
^{i}\right\rangle $ the $N$ qubit state $\left\vert \psi\right\rangle $ is
also a product state.

\section{Symmetric subspace and atomic coherent states}

Now we recall the notion of the symmetric subspace \cite{D54,ACGT72}.
Consider the standard basis, and those vectors of this basis, for which the
number of $1$-s, $N_{1}$ is fixed, and accordingly the number of $0$-s,
$N_{0}=N-N_{1}$ is also fixed. These vectors span a subspace of dimension
$\binom{N}{N_{1}}$. The number of such disjoint subspaces is $N+1$, and they
obviously exhaust the whole space. In each such subspace there is exactly one
state which is symmetric with respect of the permutations of the qubits. One
can obtain it by taking any vector $\left\vert \varphi_{N_{1}}\right\rangle $
in the given subspace, say $\left\vert \varphi_{N_{1}}\right\rangle
:=|\underbrace{0,\cdots0}_{N_{0}},$ $\underbrace{1,\cdots,1}_{N_{1}}\rangle$
(with obvious simplified notation) and then applying the symmetrization
projector $\mathcal{S}$
\begin{equation}
\mathcal{S}\left\vert \varphi_{N_{1}}\right\rangle =C\sum_{\nu}P_{\nu
}\left\vert \varphi_{N_{1}}\right\rangle , \label{sym}%
\end{equation}
where $P_{\nu}$-s are the permutation operators in $\mathbb{C}^{2^{N}}$
representing the permutation group of the qubits in the natural way, while $C$
is an appropriately chosen normalization coefficient \cite{CDL77}. Taking
all these (mutually orthogonal) states with $N_{1}=0,1\ldots N,$ they span the
$N+1$ dimensional \emph{symmetric} subspace, $\mathbb{S}$ of the total space.

We shall also need the unnormalized symmetric vectors defined as:\
\begin{equation}
\left\vert _{N_{1}}^{N}\right\rangle :=\binom{N}{N_{1}}^{\frac{1}{2}%
}\mathcal{S}\left\vert \varphi_{N_{1}}\right\rangle , \label{unnor}%
\end{equation}
which will be useful later. For example $\left\vert _{1}^{3}\right\rangle =$
$\left\vert 0,0,1\right\rangle +\left\vert 0,1,0\right\rangle +\left\vert
1,0,0\right\rangle $.

The state $\left\vert _{0}^{N}\right\rangle =|0,0\ldots0\rangle$ is obviously
symmetric and not entangled. One can consider global rotations of this latter
state by introducing the following sums of individual qubit spin operators:
\begin{equation}
J_{\alpha}=\sum_{n}^{N}S_{\alpha}^{n},\qquad(\alpha=x,y,z),\text{ }%
\end{equation}
with
\begin{equation}
S_{x}^{n}=(\left\vert 1\right\rangle \left\langle 0\right\vert +\left\vert
0\right\rangle \left\langle 1\right\vert )_{n}/2,\text{ \ }S_{y}%
^{n}=(\left\vert 1\right\rangle \left\langle 0\right\vert -\left\vert
0\right\rangle \left\langle 1\right\vert )_{n}/2i,\text{ \ }S_{z}%
^{n}=(\left\vert 1\right\rangle \left\langle 1\right\vert -\left\vert
0\right\rangle \left\langle 0\right\vert )_{n}/2.
\end{equation}
The symmetric vector $\mathcal{S}\left\vert \varphi_{N_{1}}\right\rangle $ in
(\ref{sym}) is the normalized eigenstate of $J_{z}$ with the eigenvalue
$m=(N_{1}-N_{0})/2$, and therefore we will denote it by $\left\vert
m\right\rangle $ \cite{D54,ACGT72}, where the possible values of $m$ are
$-N/2,-N/2+1,\ldots,N/2$ .

If $\mathbf{u}$ is a unit vector corresponding to a point on the unit sphere
in real three dimensional space, making the polar angle $\theta$ with the
negative $z$ axis, and the azimuth $\varphi$ with the positive $x$ axis, then%
\begin{equation}
R_{\theta\varphi}=e^{-i\theta(J_{x}\sin\varphi-J_{y}\cos\varphi)}%
\end{equation}
is a unitary rotation in $\mathbb{S}$, and the state
\begin{equation}
|\tau_{u}\rangle=R_{\theta\varphi}\left\vert _{0}^{N}\right\rangle
=R_{\theta\varphi}\left\vert m=-N/2\right\rangle .
\end{equation}
is called an atomic coherent state \cite{ACGT72}. It can be also shown that
$\left\vert \tau_{u}\right\rangle $ is the normalized eigenstate of
$\mathbf{J\cdot u}$ belonging to its highest eigenvalue $N/2$.%
\begin{equation}
\left(  \mathbf{J\cdot u}\right)  |\tau_{u}\rangle=\frac{N}{2}|\tau_{u}%
\rangle. \label{jtau}%
\end{equation}
The notation $\left\vert \tau_{u}\right\rangle $ is related to another
parametrization of the state. Following \cite{ACGT72} the vector
$\mathbf{u}$ can be alternatively charcterized by the complex number
$\tau=\tan(\theta/2)e^{-i\varphi}$, which is the stereographic projection of
the unit vector $\mathbf{u}$ to the $x$-$y$ plane. The unit vector
$\mathbf{u}$ is called sometimes the Bloch (or Poincar\'{e}) vector of this
state, similarly to the single qubit case. The state $\left\vert _{0}%
^{N}\right\rangle =|0,0\ldots0\rangle,$ which is obviously symmetric and not
entangled, is a coherent state itself, corresponding to $\mathbf{u=-\hat{z}}$
, i. e , to $\tau=0,$ and it is the eigenstate of $J_{z}$ with the eigenvalue
$m=-N/2=:-j$. \ 

\section{A state in $\mathbb{S}$ is not entangled iff it is an atomic coherent
state}

First we point out that the coherent state $|\tau_{u}\rangle$ \ is a product
state, and thus it is not entangled. \ This can be seen by \ expanding it in
terms of the symmetrized eigenstates of $J_{z}$ :%
\begin{equation}
|\tau_{u}\rangle=\sum_{m=-j}^{j}\binom{j}{j+m}^{\frac{1}{2}}\frac{\tau^{j+m}%
}{\left(  1+\left\vert \tau\right\vert ^{2}\right)  ^{j}}\left\vert
m\right\rangle ,
\end{equation}
where $J_{z}\left\vert m\right\rangle =m\left\vert m\right\rangle $. Using the
unnormalized states given in (\ref{unnor}), $\left\vert _{k}^{N}\right\rangle
$ we can write%
\begin{equation}
|\tau_{u}\rangle=\sum_{k=0}^{N}\frac{\tau^{k}}{\left(  1+\left\vert
\tau\right\vert ^{2}\right)  ^{\frac{N}{2}}}\left\vert _{k}^{N}\right\rangle
=\frac{1}{\left(  1+\left\vert \tau\right\vert ^{2}\right)  ^{\frac{N}{2}}%
}\left(  \left\vert 0\right\rangle +\tau\left\vert 1\right\rangle \right)
^{\otimes N}, \label{coh}%
\end{equation}
where we have used the binomial theorem. The latter form shows that $|\tau
_{u}\rangle$-s are product states, and therefore are not entangled.

Now we prove the reverse statement: in the totally symmetric subspace all the
nonentangled states are atomic coherent states. To this end we consider a
general linear combination of the symmetric states
\begin{equation}
\left\vert \psi\right\rangle =\sum_{k=0}^{N}\binom{N}{k}^{\frac{1}{2}}%
c_{k}\left\vert m=-\frac{N}{2}+k\right\rangle ,
\end{equation}
where the square root of the binomial coefficients have been factored out, and
the $c_{k}$-s are arbitrary numbers obeying $\sum_{k}\binom{N}{k}|c_{k}%
|^{2}=1,$ chosen to have $\langle\psi|\psi\rangle=1.$ With the unnormalized
states $\left\vert _{k}^{N}\right\rangle $ we can write%
\begin{equation}
\left\vert \psi\right\rangle =\sum_{k=0}^{N}c_{k}\left\vert _{k}%
^{N}\right\rangle ,\qquad\sum_{k}\binom{N}{k}|c_{k}|^{2}=1.
\end{equation}
The $\left\vert _{k}^{N}\right\rangle $-s have the following property
\begin{equation}
\left\vert _{k}^{N}\right\rangle =\left\vert 0\right\rangle _{n}%
\otimes\left\vert _{k}^{N-1}\right\rangle +\left\vert 1\right\rangle
_{n}\otimes\left\vert _{k-1}^{N-1}\right\rangle ,
\end{equation}
where $\left\vert _{-1}^{N-1}\right\rangle =\left\vert _{N}^{N-1}\right\rangle
=0$, by definition. The above decompositions, which correspond to the
elementary identity $\binom{N}{k}=\binom{N-1}{k}+\binom{N-1}{k-1}$ are valid
for any $n=1,\ldots N,$ as a consequence of the symmetry of the states
$\left\vert _{k}^{N}\right\rangle $ with respect of permutations. Then
\begin{equation}
\left\vert \psi\right\rangle =\sum_{k=0}^{N}c_{k}\left\vert _{k}%
^{N}\right\rangle =\sum_{k=0}^{N}c_{k}(\left\vert 0\right\rangle _{n}%
\otimes\left\vert _{k}^{N-1}\right\rangle +\left\vert 1\right\rangle
_{n}\otimes\left\vert _{k-1}^{N-1}\right\rangle ).
\end{equation}
Therefore we have
\begin{align}
l_{n}\left(  0\right)  \left\vert \psi\right\rangle  &  =\sum_{k=0}^{N-1}%
c_{k}\left\vert _{k}^{N-1}\right\rangle ,\\
l_{n}\left(  1\right)  \left\vert \psi\right\rangle  &  =\sum_{k=1}^{N}%
c_{k}\left\vert _{k-1}^{N-1}\right\rangle =\sum_{k=0}^{N-1}c_{k+1}\left\vert
_{k}^{N-1}\right\rangle .
\end{align}
\newline According to the section 1 $\left\vert \psi\right\rangle $ is not
entangled if these vectors are parallel, requiring $c_{k+1}=\tau c_{k},$ and
thus
\begin{equation}
c_{k}=\tau^{k}c_{0}%
\end{equation}
Then the nonentangled $\left\vert \psi\right\rangle $ has necessarily the
following form:
\begin{equation}
\left\vert \psi\right\rangle =\sum_{k=0}^{N}\tau^{k}\cdot c_{0}\left\vert
_{k}^{N}\right\rangle =c_{0}\sum_{k=0}^{N}\tau^{k}\left\vert _{k}%
^{N}\right\rangle =c_{0}\left(  \left\vert 0\right\rangle +\tau\left\vert
1\right\rangle \right)  ^{\otimes N}, \label{nonen}%
\end{equation}
with $c_{0}=\left(  1+\left\vert \tau\right\vert ^{2}\right)  ^{-N/2}$ as
required by the normalization condition. Comparison with (\ref{coh}) proves
the statement: in the symmetric subspace only the coherent states are nonentangled.

\section{The vectors orthogonal to $\mathbb{S}$ are all entangled}

We shall now consider vectors in $\mathbb{S}_{\bot}$ the orthogonal complement
of the symmetric space. We prove that all vectors in $\mathbb{S}_{\bot}$ are
globally entangled. Assume to the contrary, that there exists a vector
$\left\vert \varphi\right\rangle $ which can be written as a product:%
\begin{equation}
\left\vert \varphi\right\rangle =%
{\displaystyle\bigotimes\limits_{n=1}^{N}}
(a_{n}\left\vert 1\right\rangle _{n}+b_{n}\left\vert 0\right\rangle _{n}%
)\in\mathbb{S}_{\bot},
\end{equation}
with $\left\vert a_{n}\right\vert ^{2}+\left\vert b_{n}\right\vert ^{2}=1$ for
each $n.$ At least one of the $a$-s,\ say $a_{n},$ and one of the $b$-s
\ $b_{m}$ with different indices ($n\neq m$) must be zero, otherwise
$\left\vert \varphi\right\rangle $ would have a nonzero projection on
$|m=N/2\rangle=\left\vert 1,1,\ldots1\right\rangle ,$ and on $|m=-N/2\rangle
=\left\vert 0,0,\ldots0\right\rangle $ that are elements of $\mathbb{S}$,
which would contradict to $\left\vert \varphi\right\rangle \in\mathbb{S}%
_{\bot\text{.}}$. Without loss of generality we may assume that the vanishing
coefficients are $a_{1}=0$ and$\ b_{2}=0$. Then $\left\vert \varphi
\right\rangle =\left\vert 0\right\rangle _{1}\otimes\left\vert 1\right\rangle
_{2}\otimes_{n=3}^{N}(a_{n}\left\vert 1\right\rangle _{n}+b_{n}\left\vert
0\right\rangle _{n})=\left\vert 0\right\rangle _{1}\otimes\left\vert
1\right\rangle _{2}\otimes\left\vert \varphi^{\prime}\right\rangle $, where
$\left\vert \varphi^{\prime}\right\rangle $ is the remaining $N-2$ qubit state.

Now $\left\vert \varphi^{\prime}\right\rangle $\ must be orthogonal to
$\left\vert 0\right\rangle _{3}\left\vert 0\right\rangle _{4}\ldots\left\vert
0\right\rangle _{N-1}\left\vert 0\right\rangle _{N}$, as well as
to\linebreak\ $\ \left\vert 1\right\rangle _{3}\left\vert 1\right\rangle
_{4}\ldots\left\vert 1\right\rangle _{N-1}\left\vert 1\right\rangle _{N}$,
otherwise $\left\vert \varphi\right\rangle $ would contain the basis
vectors\linebreak\ $\left\vert 0\right\rangle _{1}\left\vert 1\right\rangle
_{2}\left\vert 0\right\rangle _{3}\left\vert 0\right\rangle _{4}%
\ldots\left\vert 0\right\rangle _{N-1}\left\vert 0\right\rangle _{N}$ and
$\left\vert 0\right\rangle _{1}\left\vert 1\right\rangle _{2}\left\vert
1\right\rangle _{3}\left\vert 1\right\rangle _{4}\ldots\left\vert
1\right\rangle _{N-1}\left\vert 1\right\rangle _{N}$ with some nonzero
coefficient, which means that $\left\vert \varphi\right\rangle =\left\vert
0\right\rangle _{1}\otimes\left\vert 1\right\rangle _{2}\otimes\left\vert
\varphi^{\prime}\right\rangle $\ would have a nonzero projection on
$\left\vert m=N/2-1\right\rangle $ and on $\left\vert m=-N/2+1\right\rangle .$
Therefore again, at least one of the $a_{n}$-s and the $b_{n}$-s (with
different indices) for $n=3,4,\ldots,N$ must be zero, otherwise $\left\vert
\varphi\right\rangle $ would not be in $\mathbb{S}_{\bot}.$We may set
$a_{3}=0,$ $b_{4}=0$ and continue this reasoning, until we arrive that
$\left\vert \varphi\right\rangle $ must be of the form%
\begin{equation}
\ \left\vert \varphi\right\rangle =\left\vert 0\right\rangle _{1}\left\vert
1\right\rangle _{2}\left\vert 0\right\rangle _{3}\left\vert 1\right\rangle
_{4}\ldots\left\vert 0\right\rangle _{N-1}\left\vert 1\right\rangle _{N}\text{
\qquad if}\ N\text{ is even,} \label{even}%
\end{equation}%
\begin{align}
&  a_{N}\left\vert 0\right\rangle _{1}\left\vert 1\right\rangle _{2}\left\vert
0\right\rangle _{3}\left\vert 1\right\rangle _{4}\ldots\left\vert
1\right\rangle _{N-1}\left\vert 1\right\rangle _{N}+b_{N}\left\vert
0\right\rangle _{1}\left\vert 1\right\rangle _{2}\left\vert 0\right\rangle
_{3}\left\vert 1\right\rangle _{4}\ldots\left\vert 0\right\rangle
_{N-1}\left\vert 0\right\rangle _{N}\nonumber\\
&  \text{if }N\text{ is odd.} \label{odd}%
\end{align}
In the first case (\ref{even}) has a nonzero projection on the state
$|m=0\rangle$ in the symmetric subspace, while in the second case (\ref{odd})
has a nonzero projection on the states $\left\vert m=N/2\pm1\right\rangle $
being also in the symmetric subspace.

We arrived to a contradiction: the nonentangled $\left\vert \varphi
\right\rangle $ cannot be orthogonal to $\mathbb{S}$, or stated otherwise: all
elements of $\mathbb{S}_{\bot}$ are entangled.

We thank L. Feh\'{e}r, P. F\"{o}ldi and I. Tsutsui for useful discussions. The
work was supported by the Hungarian Scientific Research Fund (OTKA)\ under
contract No: T48888.

\end{document}